\documentclass[usegraphicx,usenatbib]{mn2e}

\usepackage{amsmath}
\usepackage{amssymb}
\usepackage{color}

\title[Dark matter self-interactions in Abell 3827]{On the
  interpretation of dark matter self-interactions in Abell 3827}
\author[F.~Kahlhoefer, K.~Schmidt-Hoberg, J.~Kummer and S.~Sarkar]
  {Felix Kahlhoefer,$^1$\thanks{felix.kahlhoefer@desy.de}
  Kai Schmidt-Hoberg,$^1$ Janis Kummer$^1$ and Subir Sarkar$^{2,3}$ \\
$^1$ DESY, Notkestrasse 85, D-22607 Hamburg, Germany\\
$^2$ Rudolf Peierls Centre for Theoretical Physics, University of
  Oxford, 1 Keble Road, Oxford OX1 3NP, United Kingdom\\
$^3$ Niels Bohr Institute, Blegdamsvej 17, 2100 K{\o}benhavn {\O},
  Denmark}
\begin{document}
\maketitle
\begin{abstract}
Self-interactions of dark matter particles can potentially lead to an
observable separation between the dark matter halo and the stars of a
galaxy moving through a region of large dark matter density. Such a
separation has recently been observed in a galaxy falling into the
core of the galaxy cluster Abell 3827. We estimated the DM
self-interaction cross section needed to reproduce the observed
effects and find that the sensitivity of Abell 3827 has been
significantly overestimated in a previous study. Our corrected
estimate is $\tilde{\sigma}/m_\text{DM} \sim
3\:\text{cm}^2\:\text{g}^{-1}$ when self-interactions result in an
effective drag force and $\sigma/m_\text{DM} \sim
1.5\:\text{cm}^2\:\text{g}^{-1}$ for the case of contact interactions,
in some tension with previous upper bounds.
\end{abstract}

\section{Introduction}

Recently \cite{Massey} have studied elliptical galaxies falling into
the core of the galaxy cluster Abell 3827, following previous such
studies~\citep{2011MNRAS.415..448W, Mohammed:2014iya}. Using several
strongly-lensed images of background objects, it is possible to
reconstruct the positions of the dark matter (DM) subhaloes of each of
the galaxies. One of these is observed to be significantly separated
from the galaxy's stars by $\Delta =
1.62^{+0.47}_{-0.49}\:\text{kpc}$. While noting that an astrophysical
origin of this separation cannot be excluded, the authors interpret
this in terms of DM self-interactions and infer:
$\sigma_\text{DM}/m_\text{DM} \sim (1.7 \pm
0.7)\times10^{-4}\:\text{cm}^2\:\text{g}^{-1}$. If correct this would
rule out many attractive DM candidates, in particular axions,
neutrinos and `weakly interacting massive particles' (WIMPs) such as
supersymmetric neutralinos.

The cross section estimated by \cite{Massey} is well below the weak
upper bounds set by other astrophysical objects such as the `Bullet
Cluster' (1E 0657-56) which are typically:
$\sigma_\text{DM}/m_\text{DM} \lesssim
1\:\text{cm}^2\:\text{g}^{-1}$~\citep{Markevitch:2003at,
  Randall:2007ph, Peter:2012jh, Rocha:2012jg,
  Harvey:2015hha}. \cite{Massey} argue that A3827 is uniquely
sensitive because the infall time for the galaxies is very long, hence
the effects of DM self-interactions add up over a long period. They
adopt a simple model for the effect of DM self-interactions
following~\cite{2011MNRAS.415..448W} which predicts that the
separation between the stars and the DM subhalo should grow
proportional to $t_\text{infall}^2$ thus vastly amplifying the effect
over time.

In this letter we argue that the model used to interpret the
observations in terms of DM self-interactions is based on two
questionable assumptions:

\begin{enumerate}

\item The stars and the associated DM subhalo are assumed to develop
  completely independently such that even a tiny difference in the
  acceleration they experience can lead to sizeable differences in
  their trajectories. This neglects the crucial fact that initially
  the stars are gravitationally bound to the DM subhalo and can only
  be separated from it if external forces are comparable to the
  gravitational attraction within the system. The effect of DM
  self-interactions must therefore be at least comparable to the
  relevant gravitational effects in order to lead to an observable
  separation.

\item The effective drag force on the DM subhalo is assumed to be
  \emph{constant} throughout the evolution of the system. This is in
  contradiction with the fact that the rate of DM self-interactions
  typically depends both on the velocity of the subhalo relative to
  the cluster and the DM density of the cluster (at the position of
  the subhalo), both of which will vary along the trajectory of the
  subhalo.\footnote{The assumption of a constant drag force would be
    justified if the subhalo were on a circular orbit around the
    centre of A3827. However, the observed separation between stars
    and DM points approximately in the radial direction, so this is
    disfavoured.} In particular, the rate of DM self-interactions will
  be negligibly small as long as the subhalo is far away from the core
  of the cluster.

\end{enumerate}

These issues have previously been discussed by~\cite{Kahlhoefer}
(henceforth Ka14) in the context of merging clusters such as Abell
520, the `Bullet Cluster' or the `Musket Ball Cluster', but they apply
equally in the present context. We apply the arguments from Ka14 to
provide a corrected estimate of the DM self-interaction cross section
necessary to explain the observations of A3827.  We find values that
are intriguingly close to existing bounds from other systems, implying
that such systems can potentially be used to confirm or rule out this
interpretation.

First we provide a simple estimate of the magnitude of the drag force
and the relevant gravitational force to extract the relevant
self-interaction cross section. We refine this estimate by considering
a realistic trajectory for the subhalo and evaluating its velocity and
the background DM density along this trajectory.  We then run
numerical simulations of a subhalo falling towards the core of A3827,
accounting for the motion of a large number of DM test particles
undergoing self-interactions in a time-dependent gravitational
potential. Finally, we discuss alternatives to a simple drag force
that are motivated from a particle physics perspective.

\section{Estimates of DM self-interactions}
\label{sec:simple}

Let us assume that DM particles moving with velocity $v$ through an
ambient DM density $\rho$ experience a drag force of the form
\begin{equation}
 \frac{F_\text{drag}}{m_\text{DM}} = \frac{1}{4}
 \frac{\tilde{\sigma}}{m_\text{DM}} \, v^2 \, \rho \; ,
\end{equation}
where $\tilde{\sigma}$ is the effective DM self-interaction cross
section, $m_\text{DM}$ is the DM mass and we have chosen the
normalisation of $\tilde{\sigma}$ in such a way that the momentum
transfer cross section is given by $\sigma_\mathrm{T} =
\tilde{\sigma}/2$ in analogy to the case of isotropic scattering (see
Ka14).  Such a drag force can be obtained by averaging over a large
number of DM scattering processes with small scattering angle,
assuming that the differential cross section has no velocity
dependence but a strong angular dependence such that the probability
of scattering is peaked in the forward direction.\footnote{We note that such a cross section is very difficult to motivate from the particle physics picture. This issue will be discussed in more detail below.}  The presence of
such a drag force will slow down any DM subhalo falling into a larger
DM halo compared to objects experiencing only gravitational forces,
such as the stars bound in the subhalo.

The stars inside the subhalo, however, do not just experience the
gravitational attraction of the cluster but also the attraction of
the subhalo itself. The latter contribution aims to reduce any
separation between the subhalo and the stars and will therefore oppose
the effect of the drag force. At some point, the gravitational
attraction between DM subhalo and stars will become sufficiently large
to balance the drag force on the DM subhalo. Denoting the (average)
separation between the subhalo and the stars by $\Delta$, the
restoring force will be given by
\begin{equation}
  \frac{F_\text{sh}}{m_\text{star}} = \frac{G_\mathrm{N} \,
    M_\text{sh}(\Delta)}{\Delta^2} \; 
\end{equation}
with $G_\mathrm{N}$ Newton's constant and $M_\text{sh}(\Delta)$ the
subhalo mass within radius $\Delta$. As a rough estimate, we assume a
constant density core with radius $a_\text{sh} = 2.7 \; \text{kpc}$
and mass $M_\text{sh} = 7 \times 10^{10} \mathrm{M}_\odot$ (consistent
with the mass estimate by \cite{Massey}), so that $M_\text{sh}(\Delta)
= M_\text{sh} \Delta^3 / a_\text{sh}^3$. In order for the drag force
to result in a sizeable separation, we must require $F_\text{sh} /
m_\text{star} < F_\text{drag}/m_\text{DM}$. This inequality leads to
\begin{equation}
\frac{\tilde{\sigma}}{m_\text{DM}} > \frac{4}{v^2 \, \rho}
\frac{G_\mathrm{N} \, M_\text{sh} \, \Delta}{a_\text{sh}^3} \; .
\end{equation}
Based on our mass model for A3827 (Appendix~\ref{sec:model}), we
estimate $\rho \sim 4\:\text{GeV\:cm}^{-3}$ and $v \sim
1500\:\text{km\:s}^{-1}$ at $r = 15\:\text{kpc}$. Substituting these
values and requiring $\Delta=1.6\:\text{kpc}$ yields
\begin{equation}
\frac{\tilde{\sigma}}{m_\text{DM}} \gtrsim 2 \:\text{cm}^2\:\text{g}^{-1} \;,
\end{equation}
which is in some tension with the upper bound from the Bullet Cluster:
$\tilde{\sigma}/m_\text{DM} \lesssim
1.2\:\text{cm}^2\:\text{g}^{-1}$~(Ka14).

If the self-interaction cross section is much smaller than the
estimate obtained above, the stars will remain closely bound to the
subhalo due to the overwhelming gravitational restoring force. In
particular, it should be clear that \emph{no visible separation} can
result from a cross section as small as $\tilde{\sigma} / m_\text{DM}
\sim 10^{-4}\:\text{cm}^2\:\text{g}^{-1}$.

\subsection{One-dimensional simulations}

For the estimate above, we have assumed both the velocity of the
subhalo $v$ and the background density of the cluster $\rho$ to be
constant in time. In this case, the separation between the DM subhalo
and the stars is expected to remain constant in time, with the
equilibrium value determined by the condition
$F_\text{drag}/m_\text{DM} = F_\text{sh}/m_\text{star}$. However as
long as the subhalo is far away from the central region of the
cluster, both $v$ and $\rho$ are expected to be small, so that no
significant separation will occur. As the subhalo accelerates towards
the central region, the separation will grow, provided that the drag
force is always sufficiently larger than the restoring gravitational
force. To be more realistic, we should therefore calculate the
position and velocity of the subcluster as a function of time. To this
end, we need to also include the gravitational force of the cluster,
acting on both the subhalo and the stars:
\begin{equation}
 \frac{F_\text{cluster}}{m} = \frac{G_\mathrm{N} \,
   M_\text{cluster}(r)}{r^2} \; ,
\end{equation}
where $M_\text{cluster}(r)$ is the cluster mass within radius
$r$. Assuming a radial orbit, we can write
\begin{align}
 \ddot{r}_\text{sh} & = -\frac{F_\text{cluster}}{m_\text{DM}} +
 \frac{F_\text{drag}}{m_\text{DM}} \nonumber \\ & = -
 \frac{G_\mathrm{N} \, M_\text{cluster}(r_\text{sh})}{r_\text{sh}^2} +
 \frac{1}{4} \frac{\tilde{\sigma}}{m_\text{DM}} \, \dot{r}_\text{sh}^2
 \, \rho(r_\text{sh}) \\ \ddot{r}_\text{star} & =
 -\frac{F_\text{cluster}}{m_\text{star}} +
 \frac{F_\text{sh}}{m_\text{star}} \nonumber \\ & = -
 \frac{G_\mathrm{N} \,
   M_\text{cluster}(r_\text{star})}{r_\text{star}^2} +
 \frac{G_\mathrm{N} \, M_\text{sh}(r_\text{sh} -
   r_\text{star})}{a_\text{sh}^3} \; ,
\end{align}
where we assume the gravitational pull of the stars on the subhalo to
be negligible. As before, we will assume a constant density core for
the subhalo, such that $F_\text{sh}$ is proportional to $r_\text{sh} -
r_\text{star}$. For the cluster, however, we will use a more refined
mass model as discussed in Appendix~\ref{sec:model} and calculate the
cluster mass within radius $r$ according to $M_\text{cluster}(r) =
4\pi \int_0^r r'^2 \rho(r') \, \mathrm{d}r'$.

The set of differential equations introduced above is readily
solved for given initial conditions. We assume that the subhalo starts
falling towards the cluster from rest at an initial distance of
$100\:\text{kpc}$ and that its trajectory lies in the plane of the sky.
It will then take approximately $t_\text{infall}
\sim 10^8\:\text{yrs}$ to reach the central region of the cluster. In
the absence of a drag force, the subhalo will reach a velocity of
around $1900\:\text{km\:s}^{-1}$ at a distance of $15\:\text{kpc}$,
which is similar to the velocity assumed above.

\begin{figure}
\begin{center}
\includegraphics[width=0.78\columnwidth]{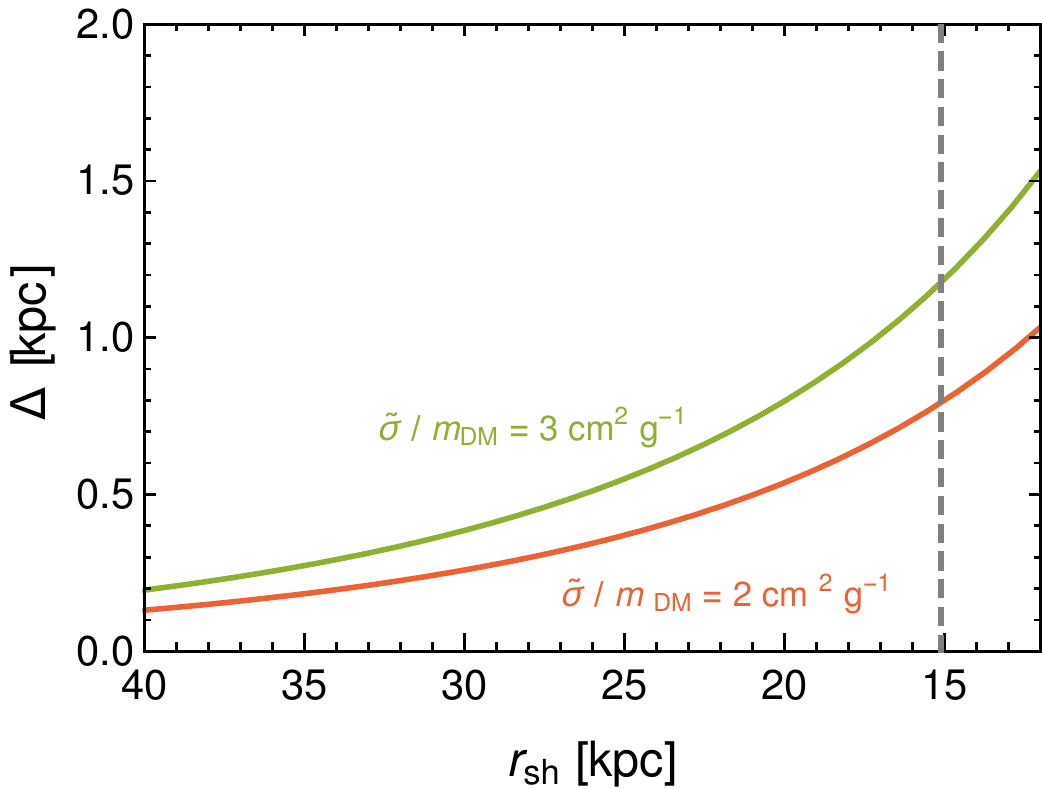}
\caption{\label{fig:1d} The separation between the DM subhalo and the
  stars as a function of the subhalo distance from the centre based on
  a simple one-dimensional simulation. The vertical line indicates the
  currently observed position of the subhalo.}
\end{center}
\end{figure}

Our results for $\Delta = r_\text{sh} - r_\text{star}$ are shown in
Fig.~\ref{fig:1d} as a function of $r_\text{sh}$ taking
$\tilde{\sigma}/m_\text{DM} = 2\:\text{cm}^2 \text{g}^{-1}$ and
$3\:\text{cm}^2 \text{g}^{-1}$. The observed separation at
$r_\text{sh} = 15\:\text{kpc}$ is found to be $\Delta =
0.8\,\text{kpc}$ and $1.2\,\text{kpc}$ respectively, showing that our
simple estimate above was quite accurate. It is worth emphasising that
the separation only becomes large once the subhalo comes close to the
central region of the cluster.

\subsection{Three-dimensional simulations}

\begin{figure*}[b]
\begin{center}
\includegraphics[width=0.79\columnwidth]{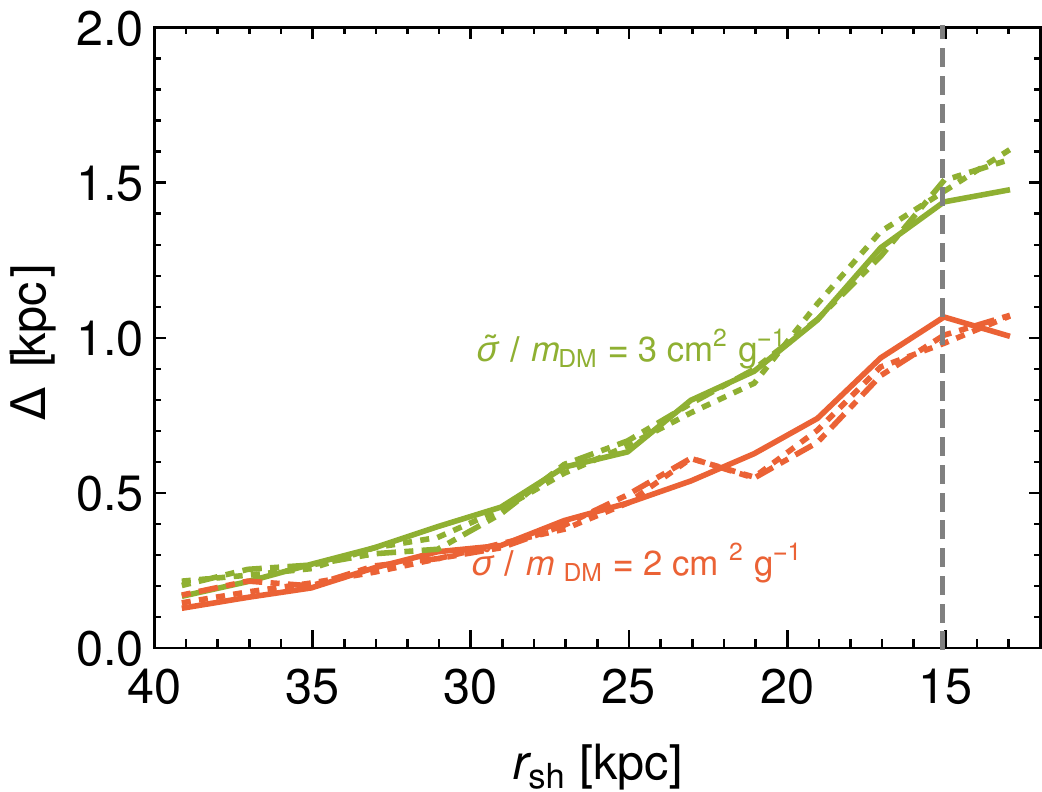}\qquad\qquad
\includegraphics[width=0.73\columnwidth]{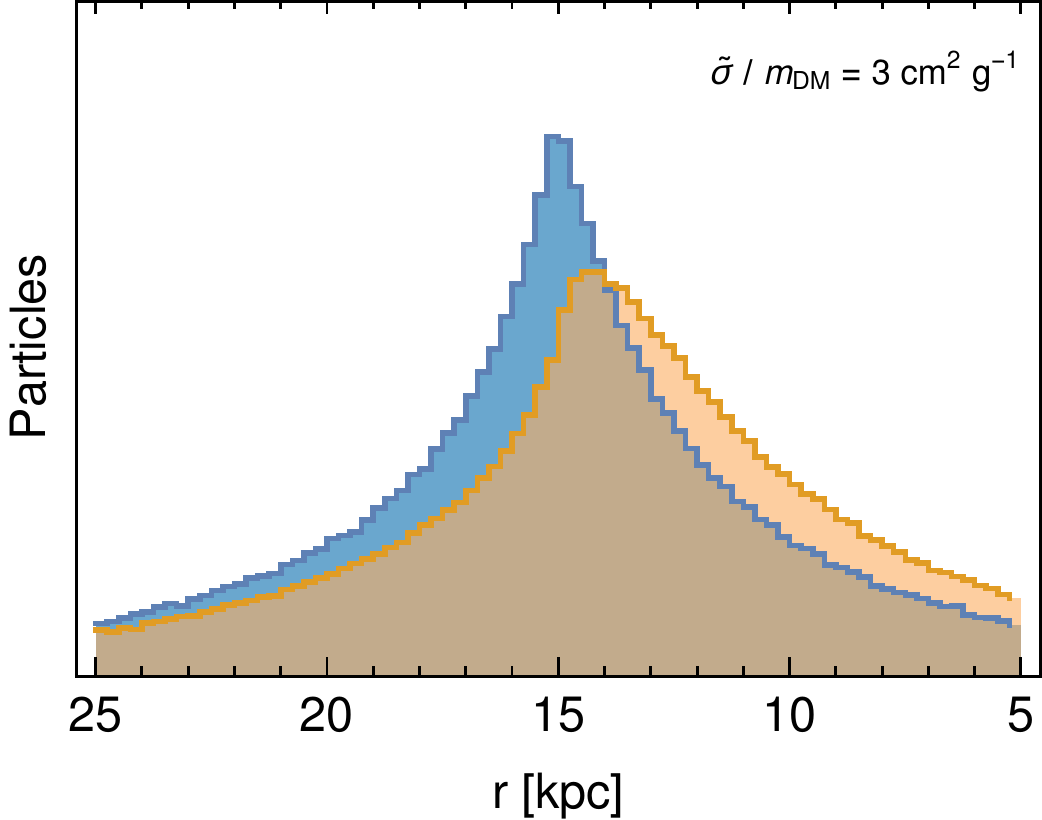}
\caption{\label{fig:3d}Left: Predicted separation between the centroid
  of the DM subhalo and the centroid of the stars resulting from an
  effective drag force on the DM subhalo as a function of the subhalo
  position $r_\text{sh}$. The solid, dashed and dot-dashed lines
  correspond to centroid calculations using the inner (20\%, 20\%),
  (5\%, 5\%) and (20\%, 5\%) of the DM subhalo and the distribution of
  stars, respectively (see text for details). Right: Histogram of the
  distribution of DM particles (blue) and stars (orange) as a function
  of radial distance $r$ from the centre of the cluster at the moment
  in time where the radial position of the peak of the subhalo is
  $r_\text{sh} = 15\:\text{kpc}$ (indicated by the vertical dashed
  line in the left panel).}
\end{center}
\end{figure*}

So far we have been using a simple model of the infalling subhalo with
a constant density core which leads to a comparably shallow potential
and a relatively small binding energy. For a more realistic halo
profile, stars in the central region will be much more tightly bound
and will therefore always remain very close to the peak of the DM mass
distribution.  Indeed, the drag force on the DM subhalo can be
interpreted as a tilt of the potential, so that loosely bound stars
are typically more affected and will preferentially travel ahead of
the DM subhalo (or even escape from the potential) and also the
minimum of the potential will be slightly shifted.  Both effects lead
to an apparent separation between the distribution of stars and the DM
subhalo.

Clearly, the motion of the stars in the combined potential of the
subhalo and the cluster cannot be modelled in a one-dimensional
simulation. In Ka14 a three-dimensional simulation was developed to
address this problem in the context of major mergers like the Bullet
Cluster. Our approach was to treat the gravitational
potential of the cluster as time-independent, while for the subhalo
the central density and scale radius are allowed to vary with time and 
determined self-consistently from the simulation. Assuming an initial
density profile, the simulation chooses a representative set of DM
particles and stars bound to the subhalo and then calculates the
motion of all these particles in the combined gravitational potential
of the cluster and the subhalo. It is then straight-forward to add in
an additional drag force affecting only the DM particles, based on the
velocity of these particles and the background DM density.

We use a \cite{Hernquist} profile to model both the cluster and the
subhalo (see Appendix~\ref{sec:model}), the advantage being that it
has a finite central potential and the velocity distribution function
can be described analytically. Since our results are quite independent
of the DM density at large radii, we expect very similar results for
the alternative Navarro-Frenk-White profile. We use the same initial
conditions for the position and velocity of the subhalo as in the
one-dimensional simulation.

A significant complication of the three-dimensional simulation is the
need to determine the centroid of the DM subhalo and the distribution
of stars. As discussed in Ka14, it is inconsistent to just calculate
these centroids including all particles which were initially bound to
the subhalo, because doing so would include particles that have
escaped from the DM subhalo and are now far away from the peak of the
distribution.  A realistic estimate of the separation can be obtained
by including only particles within the projected iso-density contour containing
20\% of the total mass of the original DM subhalo and ignoring regions
where the surface density of DM particles originating from the subhalo
is smaller than the background surface density of the cluster. This
procedure corresponds roughly to including particles within
$4\:\text{kpc}$ of the peak of the distribution. The same procedure is
applied to determine the centroid of the distribution of stars.

To illustrate how strongly our results depend on the definition of the
centroid we also show the predicted separation for different centroid
definitions. In particular, we consider the case where only the
innermost 5\% of the total mass are included in the calculation of the
centroid (such that the tails of the distribution are mostly ignored
and the position of the centroid is determined largely by the position
of the peak). While this definition may be realistic for the case of
the stars, where the peak of the distribution can be accurately
determined, it may not be appropriate for the determination of the
centroid of the DM subhalo. We therefore also consider an alternative
definition, where the inner 20\% of the DM subhalo but only the inner
5\% of the stars are included in the calculation of the centroids.

Including the tails of the DM distribution has a further important
advantage: We will see below that these tails can contain relevant
information on the nature of DM self-interactions and should therefore
not be neglected.

Our results for the separation $\Delta$ are shown in Fig.~\ref{fig:3d}
as a function of $r_\text{sh}$ for $\tilde{\sigma}/m_\text{DM} =
2\:\text{cm}^2\:\text{g}^{-1}$ and
$3\:\text{cm}^2\:\text{g}^{-1}$. The observed separation at
$r_\text{sh} = 15\:\text{kpc}$ is found to be $\Delta \sim
1.0\:\text{kpc}$ and $\sim 1.5\:\text{kpc}$, in good agreement with
our previous estimate.  We also note that the predicted separation
varies by only about 5\% as we consider different definitions of the
centroids.  This indicates that not only the tails but also the
position of the peaks differ for the two distributions.

To confirm this expectation, we show in the right panel of
Fig.~\ref{fig:3d} the distribution of the stars and the DM particles
along the radial direction. Indeed, one can see that the peaks of the
two distributions are slightly shifted. Furthermore the tail of the
distribution of stars is enhanced in the forward direction due to
stars that have escaped from the gravitational potential of the
subhalo. Most of the remaining stars, however, remain bound to the
subhalo and would return to their equilibrium position if the subhalo
survived the passage through the central region of the cluster.
\subsection{Alternative interpretations}
\label{sec:alternative}

\begin{figure*}
\begin{center}
\includegraphics[width=0.79\columnwidth]{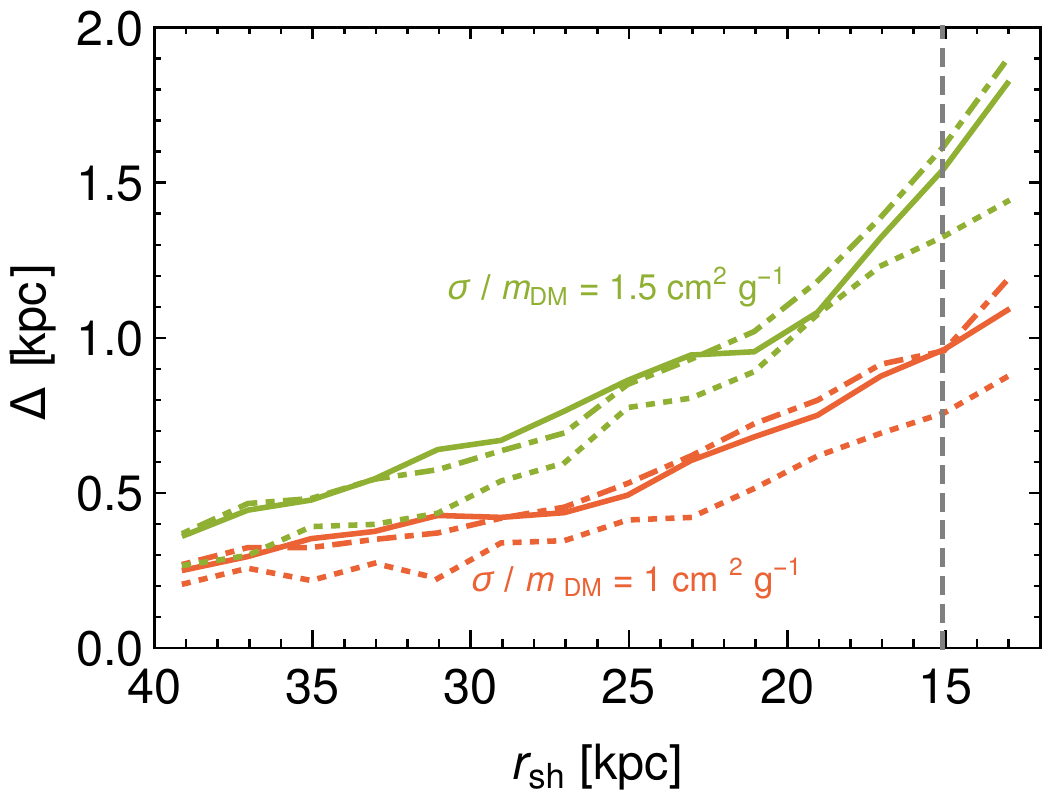}\qquad\qquad
\includegraphics[width=0.74\columnwidth]{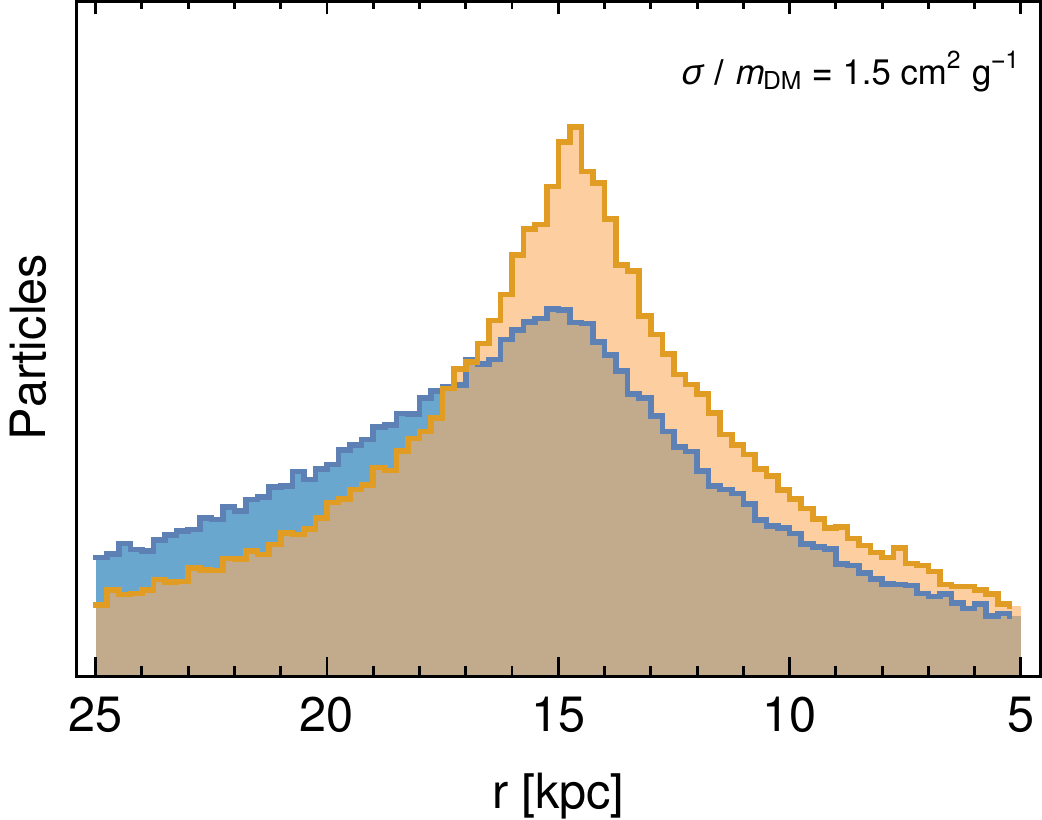}
\caption{\label{fig:rare} Same as Fig.~\ref{fig:3d} but for the case
  of rare self-interactions. In this case, the DM subhalo develops a
  tail in the backward direction, while the distribution of stars
  remains largely unchanged. The peaks of the two distributions remain
  largely coincident and therefore the separation depends sensitively
  on whether the tails of the distributions are included in the
  centroid calculation.  }
\end{center}
\end{figure*}

We briefly discuss alternative particle physics models for DM
self-interactions and their interpretation in the present context. The
two most natural forms of DM self-interactions are long-range
interactions, which can arise for example from a massless
mediator~\citep{Feng:2009mn}, and contact interactions. Long-range
interactions typically lead to a drag force of the form
\begin{equation}
 \frac{F^\text{}_\text{drag}}{m_\text{DM}} = \frac{1}{4}
 \frac{\tilde{\sigma}}{m_\text{DM}} \, \frac{c^4}{v^2} \, \rho \;
\end{equation}
with $\tilde{\sigma}/m_\text{DM} <
10^{-11}\:\text{cm}^2\:\text{g}^{-1}$ from the observation that Milky
Way satellites have survived up to the present
day~\citep{Gnedin:2000ea}. Using an estimate similar to the one
discussed above, one can immediately see that for the DM densities and
the velocities under consideration such a drag force cannot give an
observable effect. Moreover, the inverse velocity dependence implies
that DM self-interactions would actually be suppressed as the subhalo
approaches the central region of the cluster, in contrast to what is
required in order to obtain large effects from DM
self-interactions. However a massless mediator can induce
collisionless shocks, which \cite{Heikinheimo} argue can potentially
explain the features observed in A3827.

For contact interactions, an effective description in terms of a drag
force is not possible, because each individual DM particle will only
experience a small number of collisions (if any). The reason is
that in each collision the momentum transfer is so large that these
collisions must be rare to avoid observational bounds from the
survival of subhaloes and halo ellipticity~\citep{Peter:2012jh,
  Rocha:2012jg}.  Nevertheless, a separation between the DM subhalo
and stars can in principle also occur in the case of contact
interactions. In this case, the separation is not due to stars leaving
the DM subhalo in the forward direction, but due to DM particles
receiving large momentum transfer and leaving the DM subhalo in the
\emph{backward direction}. In other words, rare DM self-interactions
can affect a tail of DM particles, which shifts the centroid of the DM
distribution relative to the distribution of stars, while the peak of
the DM distribution, which is dominated by DM particles that have not
experienced any self-interactions, remains coincident with the peak of
the distribution of stars.

This scenario can also be investigated using the simulations discussed
above (for details, see Ka14). Our results are shown in
Fig.~\ref{fig:rare} for $\sigma/m_\text{DM} =
1.0\,\text{cm}^2\:\text{g}^{-1}$ and
$1.5\,\text{cm}^2\:\text{g}^{-1}$. For larger self-interaction cross
sections, the evaporation rate of the DM subhalo becomes so large that
it is unlikely to have survived up to its present position. Including
the inner 20\% of the DM subhalo and the stars in the centroid
calculation, the observed separations at $r_\text{sh} =
15\:\text{kpc}$ are found to be $\Delta = 1.0\:\text{kpc}$ and
$1.6\:\text{kpc}$, respectively. As in the case of a drag force, there
is tension between the observation of a separation in A3827 and other
constraints on the DM self-interaction cross section, which are
typically $\sigma_\text{DM}/m_\text{DM} \lesssim
1\:\text{cm}^2\:\text{g}^{-1}$~\citep{Markevitch:2003at,
  Randall:2007ph, Peter:2012jh, Rocha:2012jg, Harvey:2015hha}.

In contrast to the case of an effective drag force we find that the
predicted separation depends sensitively on the definition of the
centroids. Including only the innermost 5\% of the subhalo and the
stars (in order to determine the approximate position of the
respective peaks) yields a somewhat smaller predicted separation of
$0.8\:\text{kpc}$ and $1.3\:\text{kpc}$. This observation indicates
that the separation is mainly due to differences in the shapes of the
two respective distributions, while the peaks of the distributions
remain coincident. This expectation is confirmed by the histrograms
shown in the right panel of Fig.~\ref{fig:rare}.

An important conclusion is that the case of contact interactions can
potentially be distinguished from the case of an effective drag force
by studying in detail the shape of the DM subhalo and the relative
position of the peaks of the two distributions. In the case of contact
interactions, the DM subhalo is expected to be deformed due to the
scattered DM particles leaving the subhalo in the backward direction,
such that the position of the centroid depends sensitively on which
particles are included in the calculation. For an effective drag
force, on the other hand, we expect the DM subhalo to retain its
shape, while the distribution of stars will be both shifted and
deformed.

\section{Discussion}

In this letter we have discussed a possible interpretation in terms of
DM self-interactions of an observed separation of about
$1.6\:\text{kpc}$ between the DM halo of a galaxy and its stars. Using
several increasingly refined methods we estimate that the
self-interaction cross sections necessary to explain this effect are
of order $\tilde{\sigma}/m_\text{DM} \sim
3\:\text{cm}^2\:\text{g}^{-1}$ for the case of an effective drag force
(proportional to the square of the velocity) or $\sigma/m_\text{DM}
\sim 1.5\:\text{cm}^2\:\text{g}^{-1}$ for the case of contact
interactions. Both of these values are in tension with the upper
bounds on DM self-interactions from other astrophysical observations.

We have modelled the system under consideration rather simply, but our derived estimates are conservative since a more refined analysis (e.g.\ considering other possible trajectories for the subhalo) would predict a \emph{higher} self-interaction cross section for the same observed separation. It should therefore be
clear that A3827 is no more sensitive to DM self-interactions than
other systems considered in this context and can certainly not be used
to probe cross-sections as small as $\sigma_\text{DM}/m_\text{DM} \sim
10^{-4}\:\text{cm}^2\:\text{g}^{-1}$. Further studies of such systems
are imperative to establish if the indication from A3827 for a
non-zero self-interaction cross section (of order
$1\:\text{cm}^2\:\text{g}^{-1}$) is
indeed correct.

\subsection*{Acknowledgements}

This work is supported by the German Science Foundation (DFG) under the Collaborative Research Center (SFB) 676 Particles, Strings and the Early Universe and by the ERC starting grant ``NewAve'' (638528). SS acknowledges a DNRF Niels Bohr Professorship.

\appendix

\section{A crude model of A3827}
\label{sec:model}

For simplicity, we model both the central region of cluster A3827 and the galactic
subhalo called N1 by~\cite{2011MNRAS.415..448W} and \cite{Massey}
using a \cite{Hernquist} profile:
\begin{equation}
 \rho(r) = \frac{M}{2\pi}\frac{a}{r\,(a+r)^3} \;.
\end{equation}
For the cluster, we determine $M_\text{cluster}$ and
$a_\text{cluster}$ by fitting the mass distribution to the one given
in figure~B2 of~\cite{Massey}. This procedure leads to
$M_\text{cluster} = 7 \times 10^{13} \mathrm{M}_\odot$ and
$a_\text{cluster} = 60\:\text{kpc}$. As suggested in~\cite{Massey} we
assume a constant-density core with radius $8\:\text{kpc}$. We have
checked that this model reproduces the observed projected mass in the
central region and the observed velocity dispersion with reasonable
accuracy.

For the subhalo, we find that taking $M_\text{sh} = 5 \times 10^{11}
\, \mathrm{M}_\odot$ and $a_\text{sh} = 7\:\text{kpc}$ enables us to
reproduce satisfactorily the observed velocity dispersion and the
projected mass in the central region of the subhalo.

\end{document}